# Multi-layer $MoS_2$/GaN UV-Visible photodetector with observation of $MoS_2$ band edge in spectral responsivity

Swanand Solanke,[1] Shashwat Rathakanthiwar,[1] Anisha Kalra,[1], Muralidharan Rangarajan,[1], Srinivasan Raghavan,[1] and Digbijoy N. Nath,[1]

[1]*Centre For Nano Science And Engineering (CeNSE), Indian Institute of Science, Bengaluru, Karnataka -5600129 (INDIA)*

We report on the demonstration of $MoS_2$/GaN UV-visible photodetectors with high spectral responsivity both in UV and in visible regions as well as the observation of $MoS_2$ band-edge in spectral responsivity. Multi-layer $MoS_2$ flakes of thickness ~ 200 nm were exfoliated on epitaxial GaN-on-sapphire, followed by fabrication of detectors in a lateral Metal-Semiconductor-Metal (MSM) geometry with Ni/Au contacts which were insulated from the GaN layer underneath by $Al_2O_3$ dielectric. Devices exhibited distinct steps in spectral responsivity at 365 nm and at ~ 685 nm with a corresponding photo-to-dark current ratio of ~4000 and ~ 100 respectively. Responsivity of 0.1 A/W (at 10 V) was measured at 365 nm corresponding to GaN band edge, while the second band edge at ~ 685 nm is characterized by a spectral responsivity (SR) of ~ 33 A/W when accounted for the flake size, corresponding to the direct band gap at K point of multi-layer $MoS_2$.

There has been an increasing interest in integrating layered two-dimensional (2D) material with wide bandgap semiconductors such as group-III Nitrides due to the exciting possibilities such 2D/3D heterojunctions promise. Layered 2D materials can be deposited/transferred over almost any substrate, which allows for extreme band gap engineering, otherwise not viable using conventional heterostructures due to lattice mismatch issues[1,2]. Such 2D/3D heterojunction-based devices, which offer wide range of possibilities in terms of application-based devices, have been reported by several groups[3-6]. Many of these reports emphasize on diode like applications and report vertical current transport. However, study of these devices for optoelectronic applications is at an embryonic stage. Recently[7-10], there has been some reports on MoS$_2$-based optical detectors, but no report exists on dual-band photodetectors architecture which exploit the band gaps of both, 2D material and a wider band gap material on which such layered semiconductor is deposited or transferred. MoS$_2$ and GaN are the most widely studied 2D and wide band gap materials respectively due to their several advantages[11,12]. In this report, we demonstrate MoS$_2$-GaN heterojunction-based UV-Visible dual-band photodetectors with high spectral responsivity in both UV and visible spectral regions.

A schematic of the device structure is shown in Figure 1. GaN epi-layer was grown on sapphire using Metal-organic Chemical Vapor Deposition (MOCVD) in an Aixtron 200/4-HT horizontal flow low-pressure reactor. The GaN epilayer exhibited a smooth surface morphology with an rms roughness value of 1.5 nm (for 10 µm X 10 µm scan), measured using atomic force microscopy (AFM). The crystalline quality of the GaN sample was investigated using high resolution X-Ray Diffraction measurements. The full width at half maximum (FWHM) values of the (002) and (102) planes of GaN was measured to be 0.1 degrees and 0.3 degrees, respectively. Subsequently, the sample was subjected to standard solvent clean. MoS$_2$ flakes (obtained from bulk sample purchased from hqGraphene) were then exfoliated using standard scotch tape method. By inspection through optical microscope, multilayer (bulk) MoS$_2$ flakes were selected for device fabrication. Figure 2(a) and 2(b) revealed the flake thickness ~235 nm, thus confirming the bulk nature of MoS$_2$. Raman spectroscopy (Horiba LabRAM HR analyzer) with excitation by a 532 nm laser (Figure 2(c)) confirmed the excellent quality of flakes ($E_{2g}^1$ = 383.2, $A_{1g}$ = 408.4). Device fabrication started with e-beam lithography-based pattering for dielectric layer, followed by e-beam evaporation of 50 nm of Al$_2$O$_3$. Then, the patterning and deposition for Al$_2$O$_3$ was performed such that the; metal layer would form contacts only with MoS$_2$ and remain isolated from underlaying GaN. This was followed by patterning for metal deposition on this dielectric layer using e-beam lithography.

Ni/Au (30/100 nm) contacts were then deposited using e-beam evaporation, forming Metal-Semiconductor-Metal (MSM) (finger width of 10 µm, spacing of 10 µm and probing pad size of 150 µm square). Scanning Electron Microscope (SEM) image of device (Figure 3(a)) clearly shows that the GaN layer has been electrically isolated from the metal contacts using the dielectric. Figure 3(b) shows the top view of as-fabricated device obtained using optical microscope.

The electrical and photo-response properties of fabricated device was studied using electrical 4-probe station measurement system (Agilent B1500 Semiconductor Parameter Analyzer), Keithley 2450 source meter and ScienceTech Quantum Efficiency (QE) measurement system (Details reported previously in Ref. 13 and 14)). The QE tool was used as source of monochromatic light. Figure 4 shows the I-V characteristics of the device under dark and various illumination wavelengths. The photo-to-dark current ratio was observed to be ~ 4000 at a bias of 20 V and at an optical intensity of 8.24 mW/cm$^2$ for 365 nm. Photo-current measurements were then followed by spectral responsivity (SR) measurements using QE tool. Figure 5 shows the measured SR graph of the device from 290 nm to 1100 nm where a sharp band-edge peak (~0.04 A/W) at 365 nm corresponding to for GaN is evident. Further, we observe that the device exhibits another band-edge at ~ 685 nm (0.011 A/W), which when normalized with the active MoS$_2$ device area, corresponds to a responsivity of 33 A/W. This band-edge ideally corresponds to direct bandgap of 1.8-1.9 eV[15]. which shows that although lowest energy transition in MoS$_2$ is 1.2 eV approximately, the directly energy transition of 1.8 eV approximately still exists at *K* point. It has been reported [16,17] that, MoS$_2$ exhibits a prominent transition or peak around 1.8-1.9 eV which is attributed to direct-gap transition between split valence band maxima (*v1* and *v2*) and conduction band minima all present at *K* point of Brillouin Zone. Further, spectral responsivity exhibited a continuous 'fall-off' indicative of a layer-dependent absorption in multi-layer MoS$_2$. Figure 5 shows the spectral responsivity measured for GaN only device (without MoS$_2$) which clearly shows the absence of any band-edge above 365 nm, thus testifying that the band-edge observed at ~ 685 nm in MoS$_2$-GaN heterojunction devices was contributed from MoS$_2$. The mechanism behind obtaining such kind of behavior can be explained as follows. Figure 6(a) shows the mechanism of photo-generated carrier transport across GaN-MoS$_2$ layers. Given the thickness of MoS$_2$ is ~ 230 nm, it will absorb almost 100% of the 365nm-UV radiation incident on it, allowing negligible penetration to the GaN layer underneath. However, while the size of the MoS$_2$ flake is ~ 30 x 50 µm$^2$ and the distance between metal pads and flake edges is ~ 10 µm, the spot size of the UV illumination is ~ 2 mm in diameter. Thus, a substantial region of GaN layer outside of the MoS$_2$ flake gets exposed to the 365 nm radiation leading to generation of photo-carriers, a fraction of which are collected by the contacts on MoS$_2$ as they are swept by the fringing field, particularly those that are generated in the immediate vicinity of the MoS$_2$ flake. The small conduction band offset between MoS$_2$ and GaN is expected to play a negligible role in impeding the collection of the photo-carriers as they traverse the heterojunction.

When photon of energy equal or higher then GaN bandgap is incident on the device, generated photo-carriers will travel through MoS$_2$ and will get collected by metal pads [18-20]. The total photo current as explained in Figure 6(b) is thus given as

$$I_{Total} = I_{GaN} + I_{MoS_2} \qquad (1)$$

For wavelength <365 nm, $I_{Total}$ would be equal to Equation (1). while for wavelength greater than 365 nm $I_{GaN}$ would be zero and $I_{Total}$ will be equal to $I_{MoS_2}$.

Thus, the Spectral Responsivity (SR) can be given as,

$$R = G \frac{I_{total}}{P_{in}} \quad A/W \qquad (2)$$

Here $G$ is the internal gain, $P_{in}$ is incident optical power. Figure 7 shows the variation of peak SR with applied bias for both GaN and MoS$_2$. At 365 nm, the peak responsivity varies linearly with bias indicating transit time limited gain in GaN. At 685 nm, the responsivity saturates at higher voltage. The reduction in SR value at higher voltage can be attributed to the saturation in the magnitude of the true photo current i.e. the difference between dark current and net photo current.

Figure 8 shows the transient response for an applied bias of 20 V, indicating that the device exhibited persistent photoconductivity (PPC). Similar PPC had been reported earlier[7,21,22] for $MoS_2$-based detectors, indicating that these are slow devices. The $MoS_2$ device under study exhibited higher recombination lifetime as evident from the slow decay of the photo current on switching the light OFF, decay time $t_1$~ 2.3 seconds and $t_2$~ 26.6 seconds were extracted using a bi-exponential polynomial fit method.

The characterization of this device paves the way for new kind of architecture allowing the exploration of 2D/3D MSM-heterojunction devices for dual band photodetection applications.

In summary, we successfully demonstrated $MoS_2$-GaN dual-band photodetector using simple MSM architecture. Device showed two distinct band-edges, one for GaN at 365 nm and one at 685 nm corresponding to monolayer $MoS_2$. After 685 nm SR showed continuous fall-off till 1000 nm indicating that even though bulk nature $MOS_2$ was used, it showed band-edge corresponding to its direct band gap at K point. The present architecture can be used for large area CVD/ALD grown 2D material for exploration of dual band photodetection.

The authors would like to acknowledge support from the Research and Development work undertaken under the Visvesvaraya Ph.D. scheme of Media Lab Asia, Ministry of Electronics and Information Technology (MeitY). This work is funded by DST/SERB Grant, No. 01482. Authors would also like to thank Micro Nano Characterization Facility (MNCF) and National Nano Fabrication Centre (NNFC) staff at CeNSE, IISc for their kind support for carrying out this work.

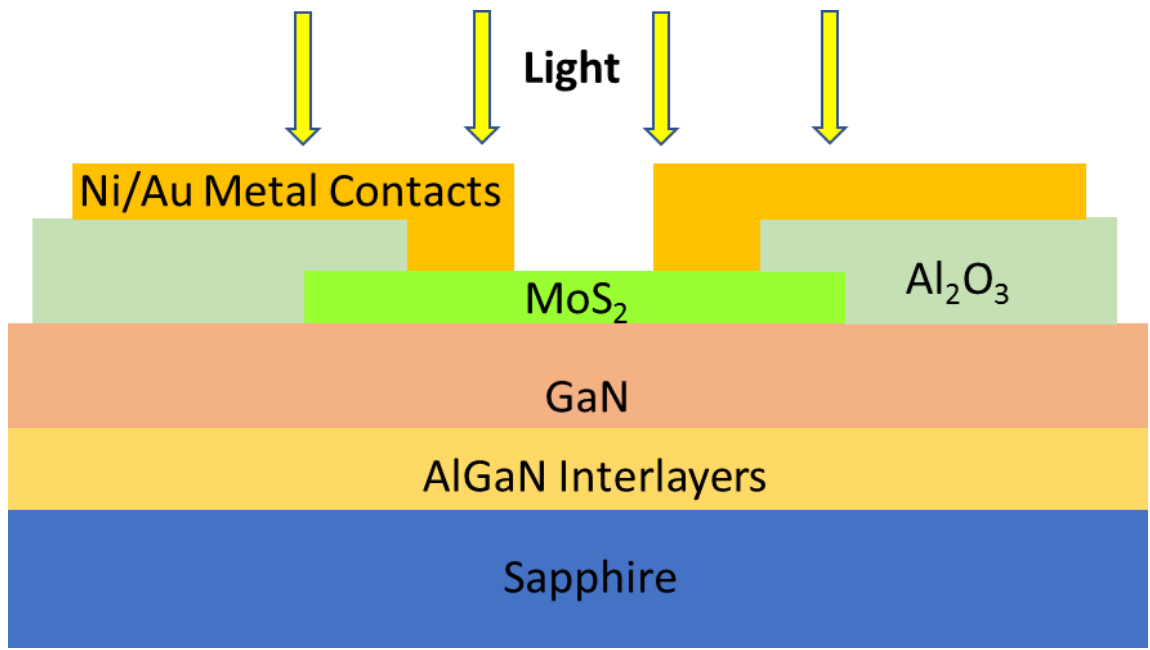

FIG. 1. Schematic of device structure showing GaN epi-stack, MoS$_2$ layer and metal pads.

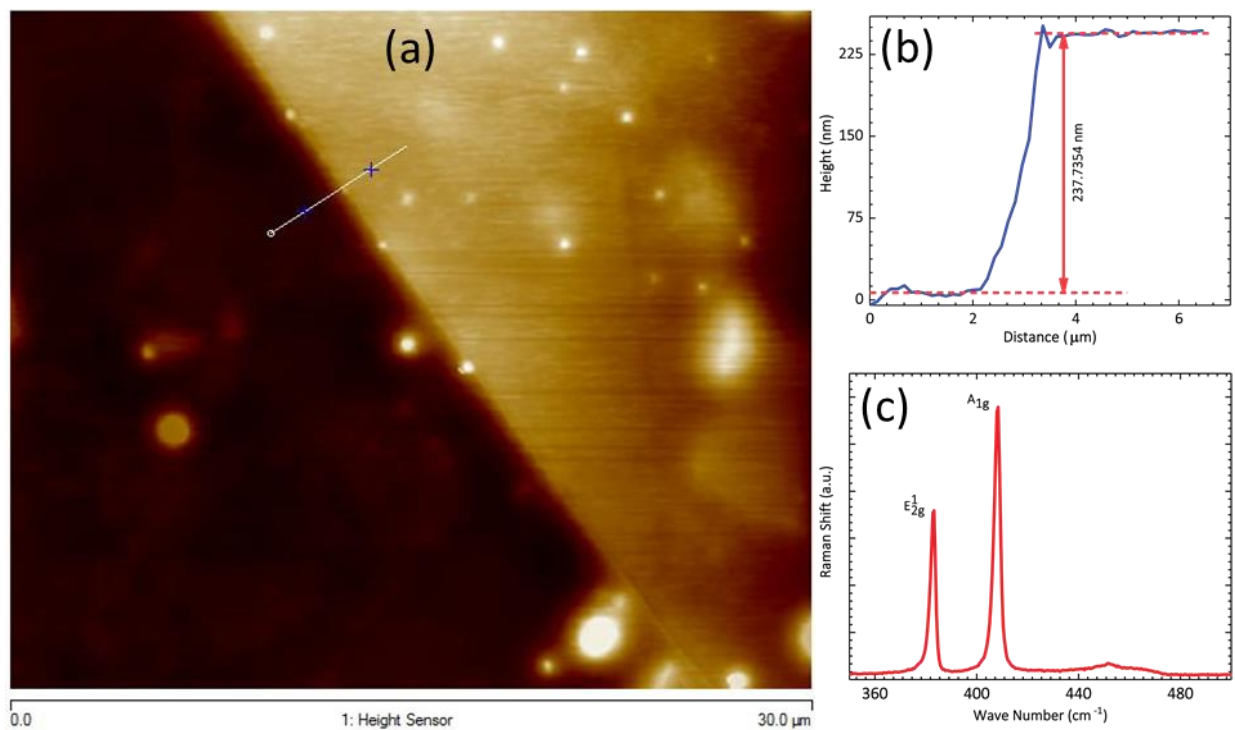

FIG. 2. (a) shows AFM image along with (b) step height measurement while (c) shows Raman spectroscopy analysis depicting bulk nature of moS$_2$.

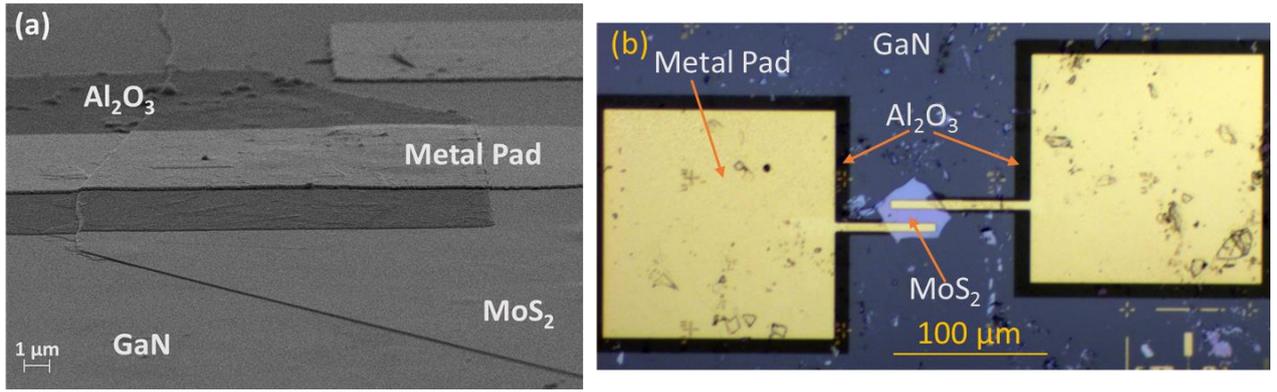

FIG. 3. (a) depicts isolation of GaN underlayer from metal pad with the help of $Al_2O_3$. (b) shows optical image of as fabricated device. Background is GaN top layer while metal pads, $Al_2O_3$ isolation layer and $MoS_2$ are shown by arrows.

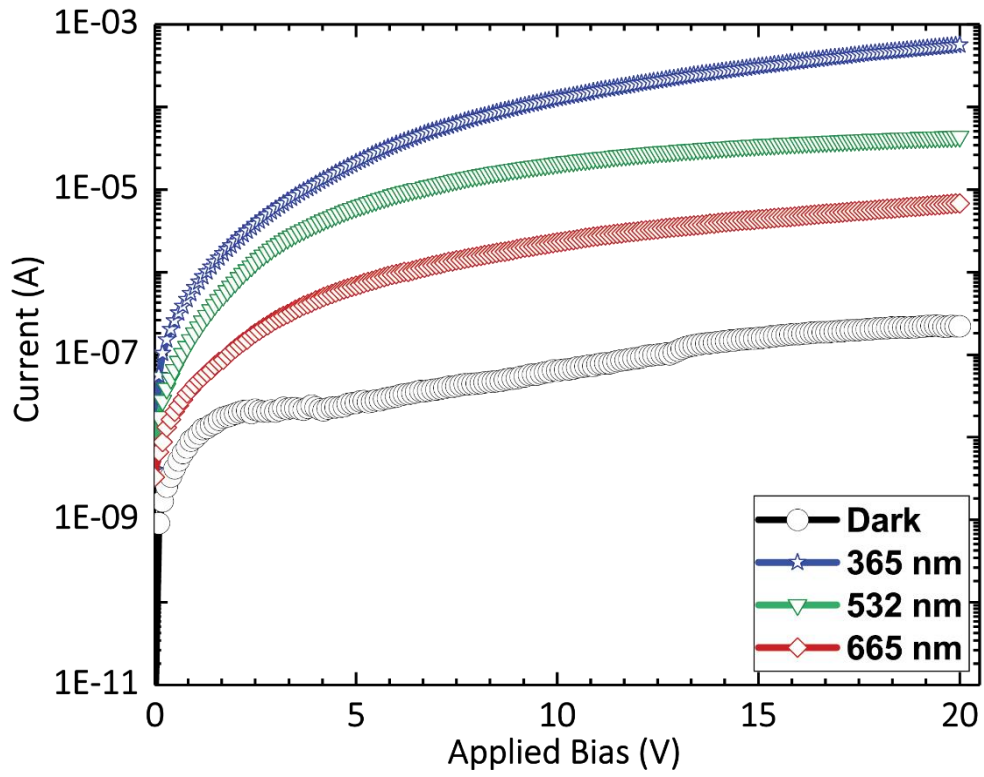

FIG. 4. Current-Voltage Characteristics (Semi-log scale) of device at select wavelengths. Measurements were carried out at room temperature.

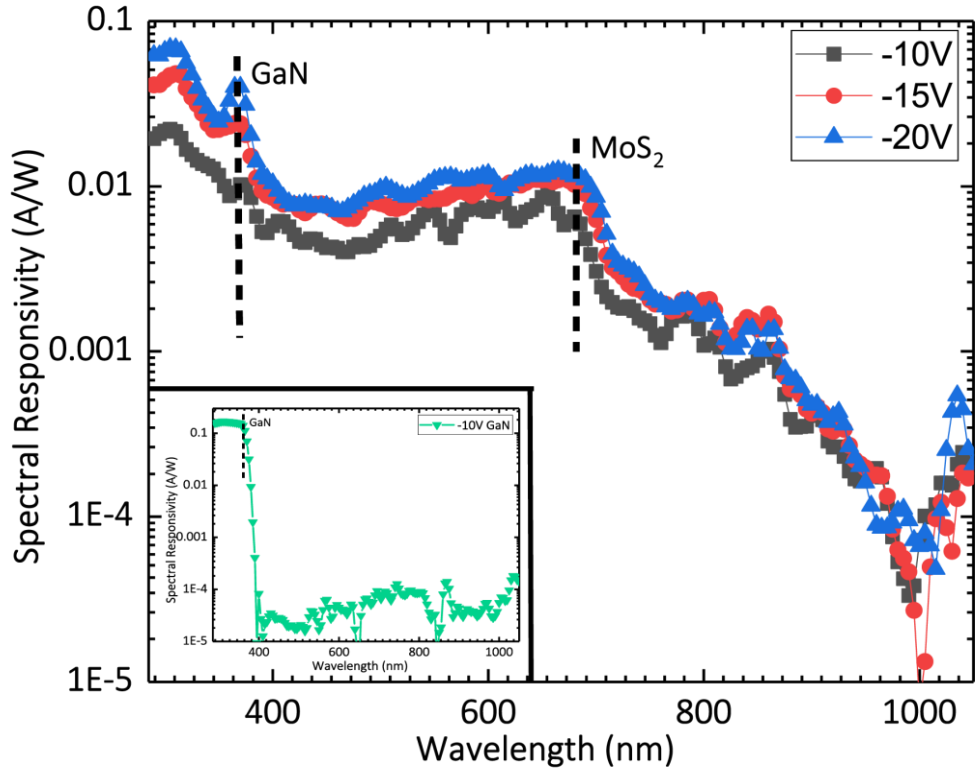

FIG. 5. Spectral Responsivity (SR) plot as a function of wavelength. SR plot for MSM device fabricated only on GaN was also shown for comparison. The absence of any band edge after 365 nm in GaN SR shows that the band edge at 685 nm in main graph has indeed shown by $MoS_2$.

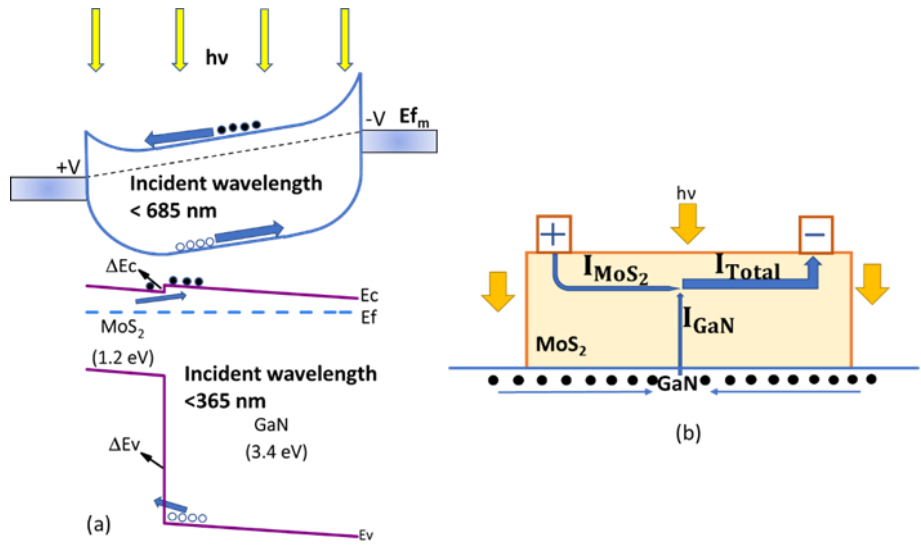

FIG. 6. (a) The Photodetection mechanism. For wavelength < 365 nm the electric field extended till GaN layer will sweep the charge carriers making them to reach to metal pads. For incident wavelength > 365 nm, GaN will not participate in photo-carrier generation and photocarriers would get generated only in $MOS_2$. (b) shows the current transport phenomenon.

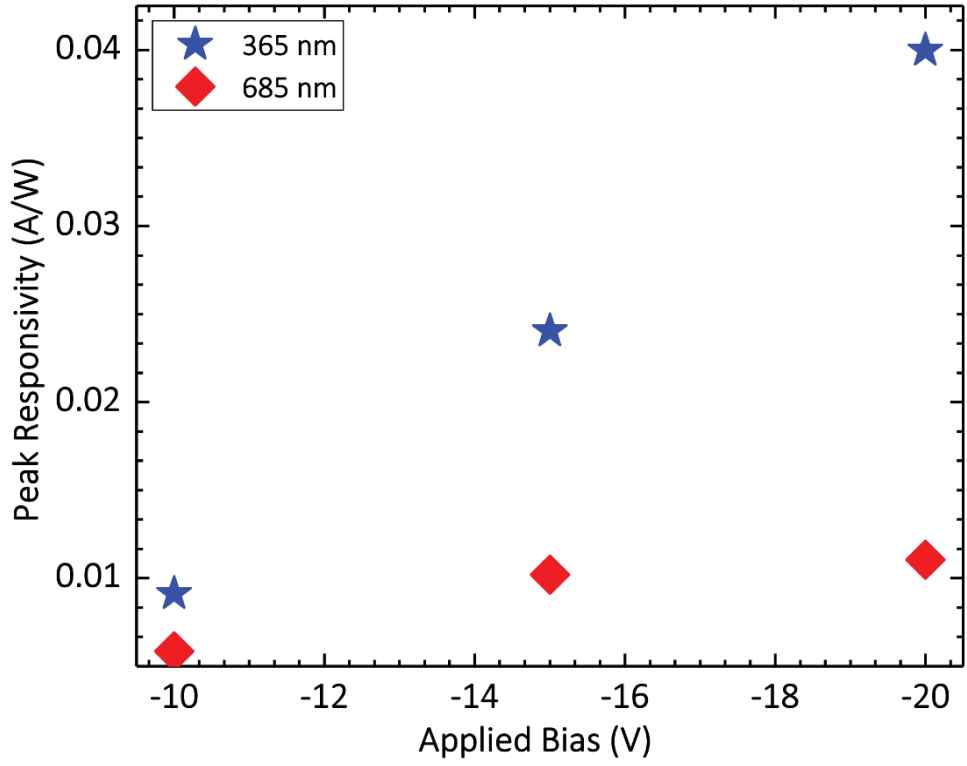

FIG. 7. Peak spectral responsivity at 365 nm and 685 nm as a function of applied bias.

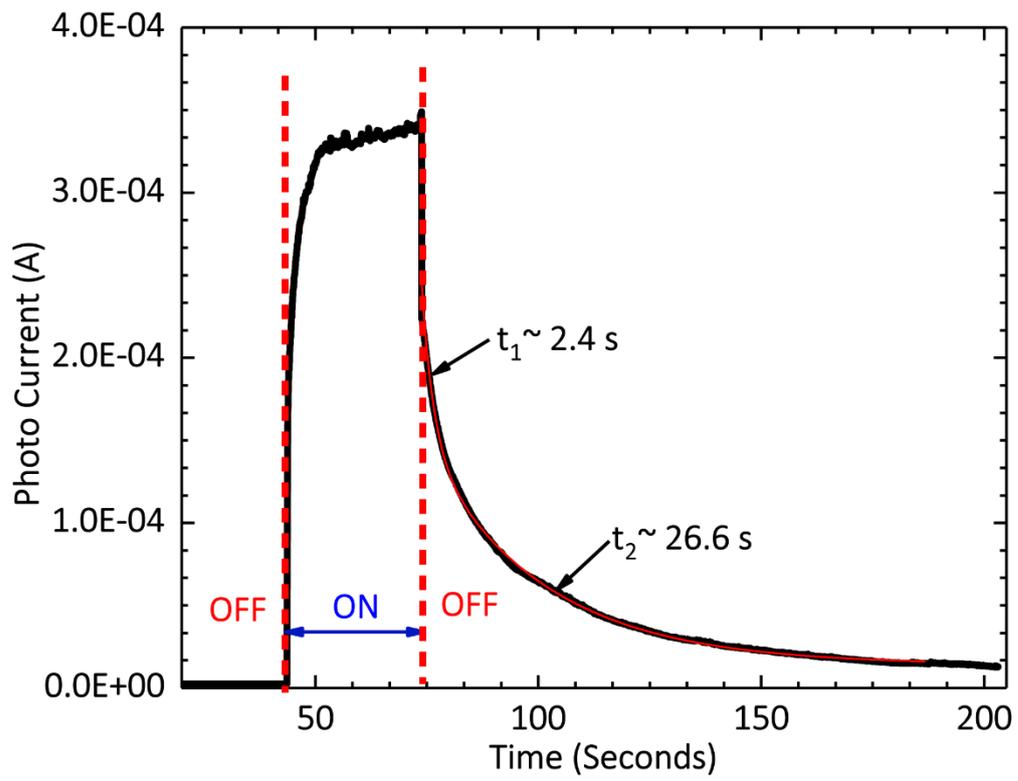

FIG. 8. Transient response of device. White light was used for measurement. Device showed a persistence photoconductivity.

15